\documentclass[prl, amsfonts, twocolumn, nofootinbib, showpacs]{revtex4}
\usepackage{graphicx}
\newcommand{\be}{\begin{equation}}
\newcommand{\ee}{\end{equation}}
\newcommand{\bea}{\begin{eqnarray}}
\newcommand{\eea}{\end{eqnarray}}

\newcommand{\gapp}{\mathrel{\raise.3ex\hbox{$>$}\mkern-14mu
              \lower0.6ex\hbox{$\sim$}}}
\newcommand{\lapp}{\mathrel{\raise.3ex\hbox{$<$}\mkern-14mu
              \lower0.6ex\hbox{$\sim$}}}


\begin{document}

\title{Difficulties for Compact Composite Object Dark Matter}

\author{Daniel T. Cumberbatch$^1$, Glenn D. Starkman$^{2}$ and Joseph Silk$^{1}$}

\email{dtc@astro.ox.ac.uk,silk@astro.ox.ac.uk,gds6@cwru.edu}

\affiliation{$^1$ Astrophysics Department, University of Oxford, Oxford, OX1 3RH, UK\\
$^2$ 
Department of Physics, Case Western Reserve University, 
Cleveland, OH~~44106-7079 
}

\begin{abstract}
It has been suggested ``that DM particles are strongly interacting composite 
macroscopically large objects ... made of well known light quarks (or ... antiquarks)."
In doing so it is argued that these compact composite objects (CCOs) are ``natural explanations 
of many observed data, such as [the] $511$ keV line from the bulge of our galaxy" observed by INTEGRAL and the excess of diffuse gamma-rays in the 1-20\,MeV band observed by COMPTEL \cite{Lawson}.
Here we argue that the atmospheres of positrons that surround CCOs composed of di-antiquark pairs in the favoured Colour-Flavour-Locked superconducting state are sufficiently dense as to stringently limit the penetration of interstellar electrons incident upon them, resulting in an extreme suppression of previously estimated rates of positronium, and hence the flux of 511~keV photons resulting from their decays, and also in the rate of direct electron-positron annihilations, which yield the MeV photons proposed to explain the 1-20~MeV excess. We also demonstrate that even if a fraction of positrons somehow penetrated to the surface of the CCOs, the extremely strong electric fields generated from the bulk antiquark matter would result in the destruction of positronium atoms long before they decay.
\end{abstract}

\maketitle

The conventional view is that dark matter is both cold -- non-relativistic --
and so weakly interacting that it is collisionless.  
Its dynamical evolution within astrophysical systems is then governed entirely by gravity.
Canonical particle physics models of dark matter
yield interactions with ordinary and dark matter 
that are indeed irrelevant at current astrophsyical densities and energies.
This conventional dark matter is therefore difficult to detect 
other than gravitationaly, despite its very considerable flux on the Earth.

Over the years, the idea that the dark matter might not be so weakly 
interacting, whether with itself or with ordinary matter, has been
explored intermittently.  In the late 1980s and early 1990s, the idea that dark matter
could carry ordinary charge was explored 
\cite{Dimopoulos:1989hk,DeRujula:1989fe,Basdevant:1989fh,Holdom}
and severely constrained.  Severe constraints were placed on the 
dark matter scattering cross section off ordinary matter 
\cite{Rich:1987st,Starkman:1990nj}.
With the exception of some windows at lower mass, 
the general conclusion was that the dark matter could not interact with ordinary
matter except weakly ($\sigma\ll10^{-30}\,{\rm cm}^2$) (and much more weakly
at typicaly WIMP masses of $10-10^3\,$GeV).  One clear exception was that
if the mass of the dark matter was sufficiently large, then the number density
of dark matter particles, and hence their flux on any natural or artificial
``detector'', would be too low to permit any useful constraints.  Thus
$\sigma_{\rm proton-DM} < 10^{-27}\,{\rm cm}^2 (m/M_{\rm Planck})$ is
unconstrained by any known astrophysical or detector limits.
(Interestingly, if $m<1$\,GeV, then most limits at high cross-section also fail
for a variety of reasons.)  Interesting generic limits are available again
only when the dark matter is sufficiently massive that its gravitational interactions
start to affect galaxy dynamics.
Spergel and Steinhardt revived the idea \cite{SIDM} of strongly self-interacting
dark matter several years ago as a way to explain the absence of central cusps in 
galaxy cores described by standard CDM cosmology \cite{CCDM_problems}.

It has been suggested \cite{Zhitnitsky2003} that the dark matter
could  be in the form of compact composite objects (CCOs hereafter) -- ``strongly interacting composite
macroscopically large objects which [are] made of well known ({\it sic}) light quarks
(or/and anti-quarks).''  Such objects, also known as QCD balls,
are ``formed from ordinary quarks [or antiquarks] during the
QCD phase transition when ({\it sic}) axion domain walls undergo
an unchecked collapse due to the surface tension which exists in the wall'' \cite{Zhitnitsky2003}.
An important  prediction of \cite{Zhitnitsky2003} is that the minimum baryon number of
a CCO for which its internal Fermi pressure renders it absolutely stable against the surface tension in the axion domain wall surrounding it during formation is $B_{\rm CCO}\simeq10^{33}$. However, metastable CCOs may form with Baryon numbers as small as $10^{20}$, based on limits from the non-detection of neutral soliton-like objects by the Gyrlyanda experiments at Lake Baikal \cite{Zhitnitsky_review}. 

With the mass of the CCO $M_{\rm CCO}\sim B_{\rm CCO}\,$GeV, traditional limits on strongly interacting massive particles
\cite{Starkman:1990nj} are clearly not applicable -- the flux of CCOs is far too low
to register in any detector, including most astrophysical ones.  For $B_{\rm CCO}=10^{33}$,
CCOs impact the Earth a few times per year, the Sun perhaps $10^4$ times per year,
and a typical neutron star, just once every $10^5$ years.  
Individual impact events on Earth might be visible to cosmic ray detectors, 
but the instrumented area of the Earth  is far too small. However, it has been highlighted that 
seismic shock waves resulting from the passage of a CCO
through the Earth may already have been detected \cite{Zhitnitsky2003}.

Individual impacts of CCOs with the Sun are unlikely to be observable.
Because the CCO bulk matter consists of extremely dense superconducting antiquarks/quarks, the majority of hadrons traversing the interstellar medium possess kinetic energies far below the superconducting gap and are therefore unlikely to penetrate
the CCO before being elastically scattered.
The geometric cross-section of a CCO is given by
\be
\sigma_{\rm CCO}=\pi R_{\rm CCO}^2.
\ee

\noindent where the CCO radius $R_{\rm CCO}$, is determined by equating the Fermi pressure in the bulk matter with the pressure associated with the surface tension in the domain wall forming it. Follwing the treatmnent in \cite{Zhitnitsky2003}, the typical radius is then given by

\be
R_{\rm CCO}\simeq\left(\frac{c}{8\pi\sigma}\right)^{1/3}B_{\rm CCO}^{4/9},
\ee

\noindent where $c\sim0.7$ is related to the degeneracy associated with the massless degrees of freedom of each CCO and $\sigma\sim10^{8-12}$\,GeV$^{3}$ is the axion domain wall tension, which is constrained by axion search experiments. 
A CCO with $B_{\rm CCO}=10^{33}$ will therefore possess a typical radius $R_{\rm CCO}\sim(10-100)\mu\,m$, and strike
 
\be
N_{\odot} = \sigma_{\rm CCO} {\rho_{\odot}\over {\rm GeV}} 2R_{\odot} \simeq 10^{29-31}
\ee

\noindent nucleons on its way through the Sun, transferring approximately $10^{23-25}$\,GeV
of energy, but only at a rate of $10^{12-14}$\,GeV\,cm$^{-1}$ , or $10^{17-19}$erg\,s$^{-1}$. 
This is $10^{-(14-16)}$ of the solar luminosity.
Nevertheless, this does suggest that if $B_{\rm CCO}\leq10^{33}$ then
CCOs will be significantly slowed down, and therefore captured within the Sun.
At $B_{\rm CCO}\simeq10^{33}$, a maximum of $\sim10^{12}$ CCOs could have been captured by the Sun to date.
It is difficult to see how to argue generically that the Sun has not captured
this number of CCOs, since $10^{44}$ baryons represents less than $10^{-13}$ of the 
Sun's baryon number.   Furthermore, one expects the CCOs, being so heavy,
if stable inside the solar environment, 
will settle to the centre of the sun.  If the CCOs were point particles,
they might form a black hole which would consume the sun \cite{Starkman:1990nj},
but they are composite objects of much too 
low a density to form a black hole of such low mass.  
Instead, they are likely to either dissolve in the Sun,
or combine into a single QCD ball of increasing size, until
they reach the maximum stable mass, after which additional 
CCOs will indeed dissolve.

Similar considerations apply to CCOs that strike neutron stars.
Individual impacts while reasonably energetic, are not sufficiently
so to be observed across the galaxy -- even the complete
annihilation of a $B=10^{33}$ CCO releases only $10^{30}$\,ergs,
typically over a (significant) fraction of a second. 
(The luminosity of the Sun, by comparison,
is $4\times10^{33}$\,erg\,s$^{-1}$.)  
However, much of the energy release occurs deep within the neutron star,
so that the most of the energy eventually goes into heating the neutron star.
The rise in temperature of the neutron star as a result of such an
event is less than $1$ degree.

One of the claimed attractions of CCO dark matter is that it could potentially explain several astronomical observations which indirectly indicate the possible presence of dark matter. Chief among these are the 511~keV line signal from the centre of the galaxy and the apparent excess in background gamma-rays of energies approximately in the range 1-20~MeV.

Firstly, we consider the 511\,keV gamma-ray line from the galactic centre measured using the SPI spectrometer of the INTEGRAL satellite \cite{Teegarden}.  This annihilation signal is thought to be produced by the process
\be
e^+ e^- \to 2\gamma.
\ee

\noindent Spectral analyses of the line signal strongly indicate that such annihilations occur following the decay of positronium atoms (25\% as para-positronium and 75\% as ortho-positronium) consisting of electrons and positrons traversing the interstellar medium (ISM) \cite{jean}. It has been proposed that such atoms may form when low energy electrons traversing the ISM, interact with low energy positrons contained within the atmosphere of positrons (or positronsphere) predicted to surround each antimatter CCO \cite{Oaknin, Zhitnitsky}, since for CCOs in the favoured colour-flavour-locked (CFL) superconducting phase, leptons are prohibited from entering the bulk matter (see e.g.~\cite{Rajagopal}). Each positronsphere is electrostatically bound to an antimatter CCO by its extremely large net negative charge 

\be
Q_{\rm CCO}\simeq-0.3eB_{\rm CCO}^{2/3},
\label{Q}
\ee

\noindent resulting from a deficiency of strange antiquarks in a thin surface layer of the CCO bulk matter, owing to its finite surface area \cite{Madsen}. The positronium atoms are then thought to survive until they decay, upon which they emit two 511~keV photons, which are claimed to contribute to the observed line signal. 

In \cite{Zhitnitsky}, it is claimed that the observed rate of 511~keV photon production is ``not in contradictions withobservations for sufficiently large $B_{\rm CCO}$''. This is obviously true since the annihiliation rate in the above scenario per unit volume per unit time, as a function of the distance, $r$, from the galactic centre is given by eq.~(2) of \cite{Oaknin} as

\begin{eqnarray}
\frac{dW}{dV dt} (r) &\simeq& 4\pi R_{\rm CCO}^2~v_{\rm rel.}~n_B(r)~n_{\rm DM}(r)\nonumber \\
&=& 4\pi R_{\rm CCO}^2~v_{\rm rel.}~\frac{\rho_B(r)}{1 {\rm GeV}}\frac{\rho_{\rm DM}(r)}{B_{\rm CCO}~1{\rm GeV}} \propto B_{\rm CCO}^{-1/3},\nonumber\\
\label{rate1}
\end{eqnarray}

\noindent where $\rho_B, n_B$ are the energy and number densities of galactic baryons repectively, $\rho_{\rm DM}$ is the energy density of dark matter particles (which in the present context are CCOs), and $v_{\rm rel.}$ is the relative speed between colliding baryons and CCOs. Thus, we observe that a sufficiently large value of $B_{\rm CCO}$ will yield an annihilation rate consistent with the observations by INTEGRAL.

However, the rate (\ref{rate1}) assumes that {\em every} electron incident on a CCO will form positronium and ignores the relative suppression owing to the fact that such electrons will only be able to interact with the lower density regions of the positronsphere due to the increasing levels of electrostatic repulsion as it penetrates further into the positronsphere. In what follows we estimate the value of the suppression factor $P$ associated with this effect. But firstly we must have a description of the electrical properties of the positronsphere surrounding each CCO. To do so, we follow the treatment by Hu and Xu in \cite{Hu}, which is largely influenced by the original calculations relating to strange stars by Alcock, Fahri and Olinto \cite{Alcock}.

The standard calculation proceeds by assuming that the quarks and positrons near the surface of the CCO bulk matter are locally in thermal equilibrium; the relationship between the charge density and electric potential is described by the classical Poisson equation. The thermodynamic potentials $\Omega_i$ as functions of the chemical potentials $\mu_i$ (where $i=\bar{u}$,$\bar{d}$,$\bar{s}$,$e^+$ for up, down and strange antiquarks and positrons respectively), the strange quark mass $m_s$ and the strong coupling constant $\alpha_c$ can be found in the literature \cite{Alcock}. Chemical equilibrium between the weak interactions involving the three antiquark flavours and positrons is maintained by the conditions

\begin{eqnarray}
&&\mu_{\bar{d}}=\mu_{\bar{d}}=\mu,\\
&&\mu_{e^+}=\mu_{\bar{u}}=\mu,
\end{eqnarray}

\noindent and their number densities are determined and related by the relations

\begin{eqnarray}
&&n_i=-\frac{\partial\Omega_i}{\partial\mu_i},\label{n_i}\\
&&\frac{{\rm d}^2V}{{\rm d}z^2}=n_{e^+}+\frac{1}{3}n_{\bar{d}}+\frac{1}{3}n_{\bar{s}}-\frac{2}{3}n_{\bar{u}},
\end{eqnarray}

\noindent where $z$ is the altitude above the surface of the surface of the bulk matter.

The quark number densities drop to zero for $z>0$, but are not uniform for $z<0$ when one correctly accounts for the finite surface area of the CCO. This results in a significant depletion of $\bar{s}$ in a thin surface layer of the bulk matter (with a thickness of order $\alpha_c^{-1}\sim1~$fm), resulting in a net negative charge of the bulk matter $Q_{\rm CCO}$, given by equation (\ref{Q}) \cite{Madsen}. The chemical potential $\mu$ can then be determined by ensuring that the surface pressure of the bulk matter is zero. 

In the widely-adopted Thomas-Fermi model, the positrons are approximated by a non-interacting Fermi gas, with the Fermi momentum $p_F$ of the positrons at each altitude being equal to the electric potential $V$. The positron number density is then given by

\begin{equation}
n_{e^+}(z)=\frac{V^3}{3\pi^2}, \quad{\rm for}~z>0.
\label{n_e}
\end{equation}

Note that the thermodynamical potentials $\Omega_i$ are defined for $V=0$ and for zero temperatures. This is corrected by  substituting $\mu_i\rightarrow\mu_i-q_iV$ in (\ref{n_i}), where $q$ are the respective electric charges \cite{Hu}, and by adopting the reasonable scenario where the temperature of the CCOs is much less than the potential energy at $z=0$ \cite{Usov_fields, Alcock}. Adopting such a scenario yields

\begin{equation}
\label{a6}
\frac{{\rm d}^2V}{{\rm d}z^2}=\left\{\begin{array}{l l}
{\displaystyle \frac{V^3}{3\pi^2}+\frac{1}{3}n_{\rm
s}(V)+\frac{1}{\pi^2}\Big[\frac{1}{3}\Big(\mu+\frac{1}{3}V\Big)^3}&\\[10pt]
~~~{\displaystyle -\frac{2}{3}\Big(\mu-\frac{2}{3}V\Big)^3
\Big]\Big(1-\frac{2\alpha_{\rm c}}{\pi}\Big)},& z<0\\[10pt]
{\displaystyle \frac{V^3}{3\pi ^2},}& z\geq 0
\end{array}\right\}
\end{equation}\\

where the complex term $n_{\rm s}(V)$ can be derived from the chemical potentials $\Omega_i$, but we omit here for clarity.

The boundary conditions for Eq.(\ref{a6}) are
\begin{eqnarray}
z&\rightarrow&-\infty:\quad V\rightarrow V_0,\quad {\rm d}V/{\rm d}z\rightarrow0,\nonumber\\
z&\rightarrow&+\infty:\quad V\rightarrow 0,\quad~~ {\rm d}V/{\rm d}z\rightarrow0,\nonumber\\
\label{boundary_conditions}
\end{eqnarray}

\noindent where $V_0$ is the electric potential
in the deep core of the CCO, and for present purposes, has a value
that is almost indiscernable from that of the surface potential $V_{\rm c}$.

By integrating Eq.(\ref{a6}) in $(-\infty,V_{\rm c}]$ and $[V_{\rm
c},+\infty)$ respectively, and invoking the boundary conditions (\ref{boundary_conditions}), 
corresponding expressions for the electric field $E=-{\rm d}V/{\rm d}z$ are obtained. 
The surface potential $V_{\rm c}=V(z=0)$ is clearly a function of
$\alpha_{\rm c}$, but here we adopt the conventional value of
$V_c=20~$MeV, used widely throughout the relevant liteature (see e.g. \cite{Lawson, Alcock}), and 
obtained by substituting the canonical CCO bulk matter 
density $n_q(0)=n_e(0)\sim9n_0$ into eq.~(\ref{n_e}), where $n_0=0.15~$fm$^{-3}$ is the characteristic 
nuclear density.

Substituting $V_{\rm c}$ into the equations for $E$ and
then integrating them, expressions describing the electrical properties of the CCO positronsphere for $z\geq0$ are
\footnote{
We should note that the 1-D approximations eq.(\ref{V})-(\ref{n_p}) are particularly appropriate for altitudes $z<R_{\rm CCO}$, and despite the fact that we utilise them in our calculations at larger altitudes it is unlikely that the deviations from an exact 3-D treatment would significantly alter our conclusions.
}
\begin{equation}\label{V}
{\displaystyle V(z)=\frac{V_{\rm c}}{1+{\displaystyle \frac{V_{\rm
c}z}{\sqrt{6}\pi}}},}
\end{equation}
\begin{equation}\label{E}
{\displaystyle E(z)=-\frac{{\rm d}V}{{\rm
d}z}=\frac{1}{\sqrt{6}\pi}\frac{V_{\rm c}^2}{\Big
(1+{\displaystyle \frac{V_{\rm c}z}{\sqrt{6}\pi}}\Big )^2},}
\end{equation}
\begin{equation}\label{n_p}
{\displaystyle n_{\rm
e}^{+}(z)=\frac{V^3}{3\pi^2}=\frac{1}{3\pi^2}\frac{V_{\rm c}^3}{\Big
(1+{\displaystyle \frac{V_cz}{\sqrt{6}\pi}}\Big )^3}.}
\end{equation}

We can utilise equation (\ref{E}) to calculate the approximate trajectory of an incident electron traveling along a radial vector using the classical equation of motion

\begin{equation}
\frac{{\rm d}^2z}{{\rm d}t^2}=\frac{eE(z)}{m_e}.
\end{equation}

\begin{figure}
     \begin{center}
     \includegraphics[width=90mm, height=50mm, clip]{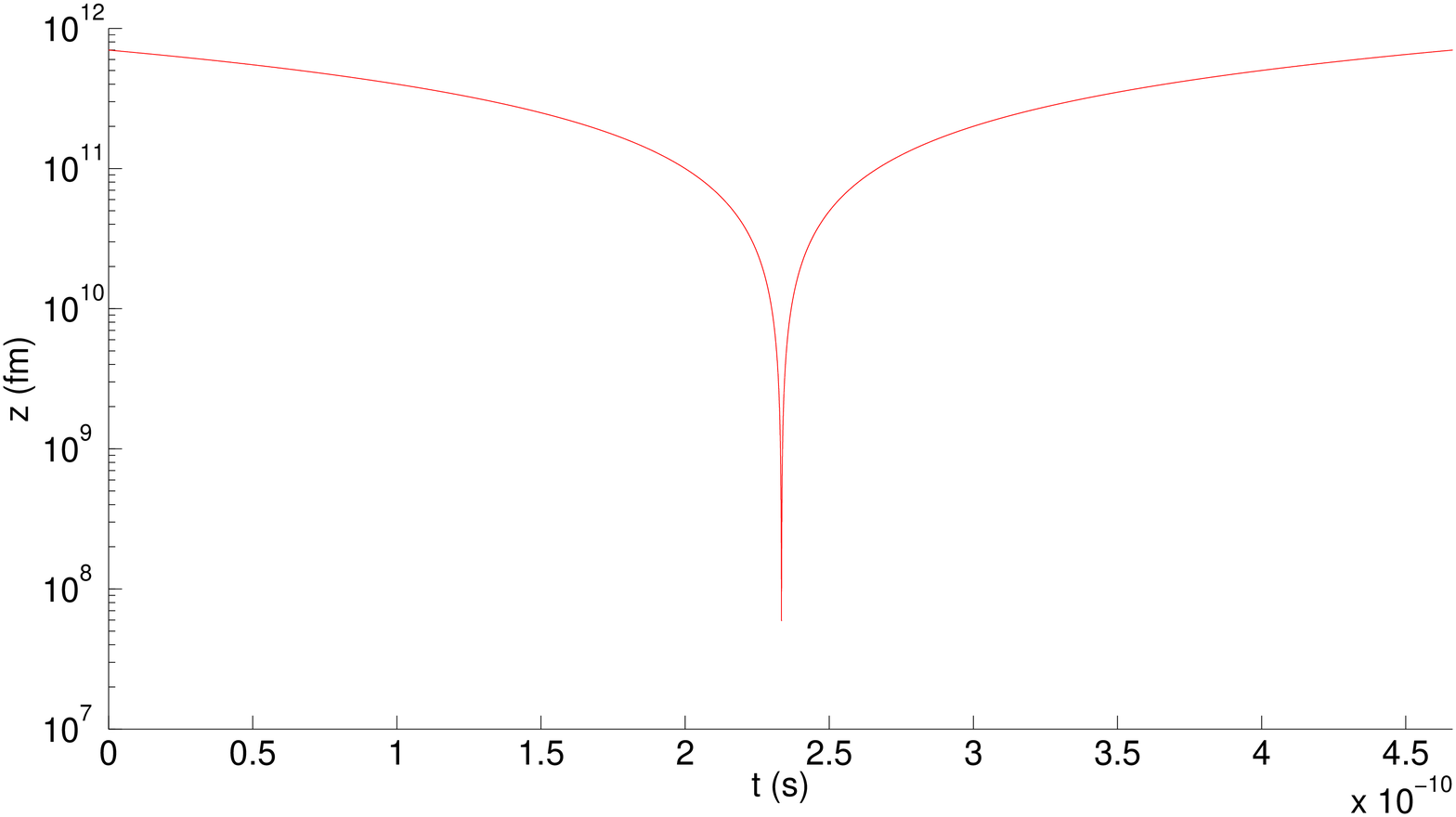}
     \includegraphics[width=90mm,height=50mm,clip]{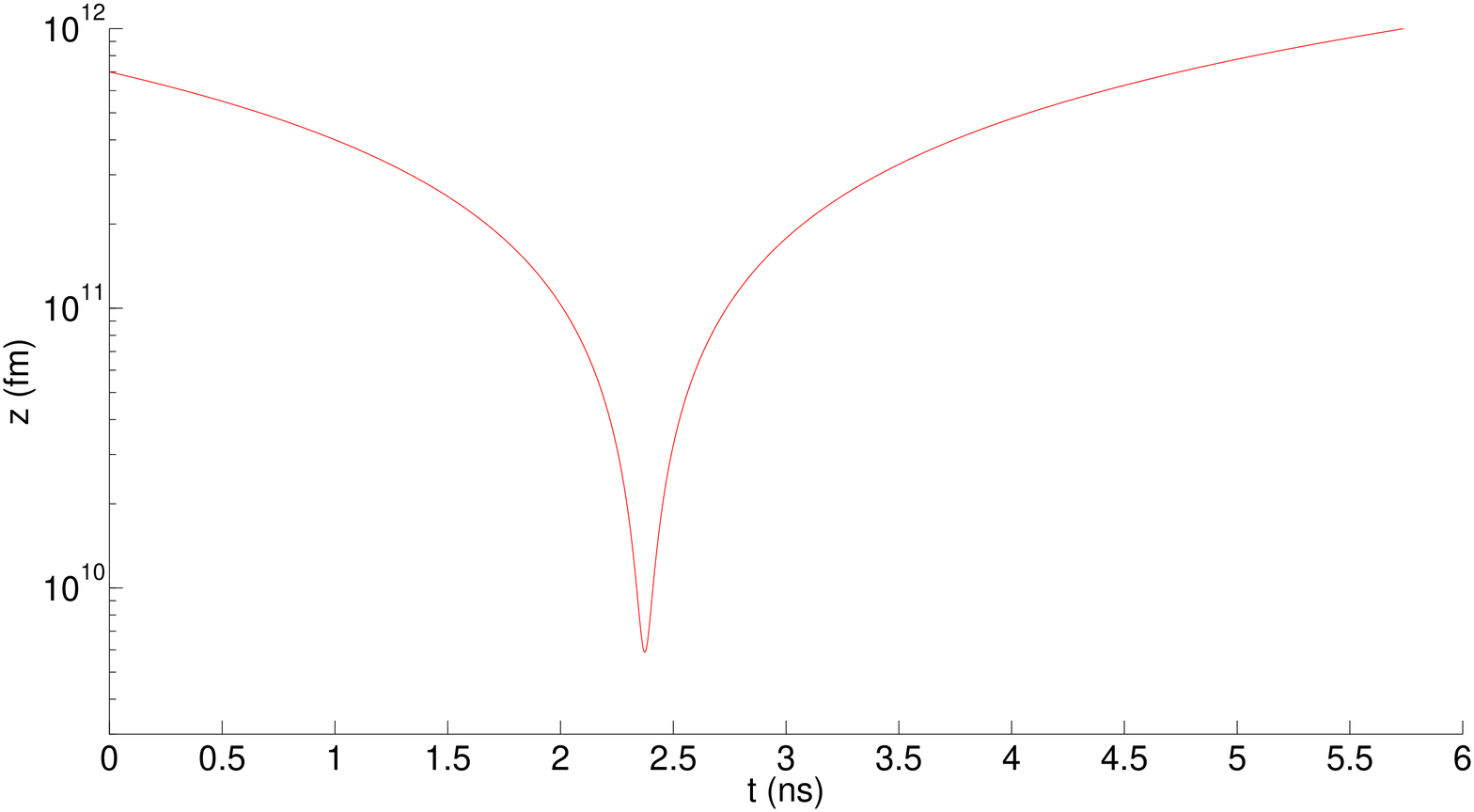}
     \end{center}
     \vspace{-0.5cm}
     \caption{Classical trajectory for an electron incident with the positronsphere of a CCO with an initial relative speed $\beta_{e^-}^{\rm ini.}$, equal to $10^{-2}$ (top) and $10^{-3}$ (bottom).}
     \label{classical_trajectory}
     \vspace{-0.3cm}
\end{figure}

Figure~\ref{classical_trajectory} displays the classical trajectory of an electron approaching a CCO from $z=\infty$ with initial speeds $\beta_{e^-}^{\rm ~ini.}=10^{-3}$ and $10^{-2}$, which are the approximate extremes of the range of speeds expected for the majority of elctrons traversing the ISM, given the rotational velocity of dark matter and baryons in the galaxy. 
Clearly, we observe that such electrons only interact with positrons at altitudes $z\geq6~\mu\rm{m}$, which in fact, as mentioned above, is of the same order as the typical radius of the CCO itself.

\begin{figure}
     \begin{center}
     \includegraphics[width=90mm,height=50mm,clip]{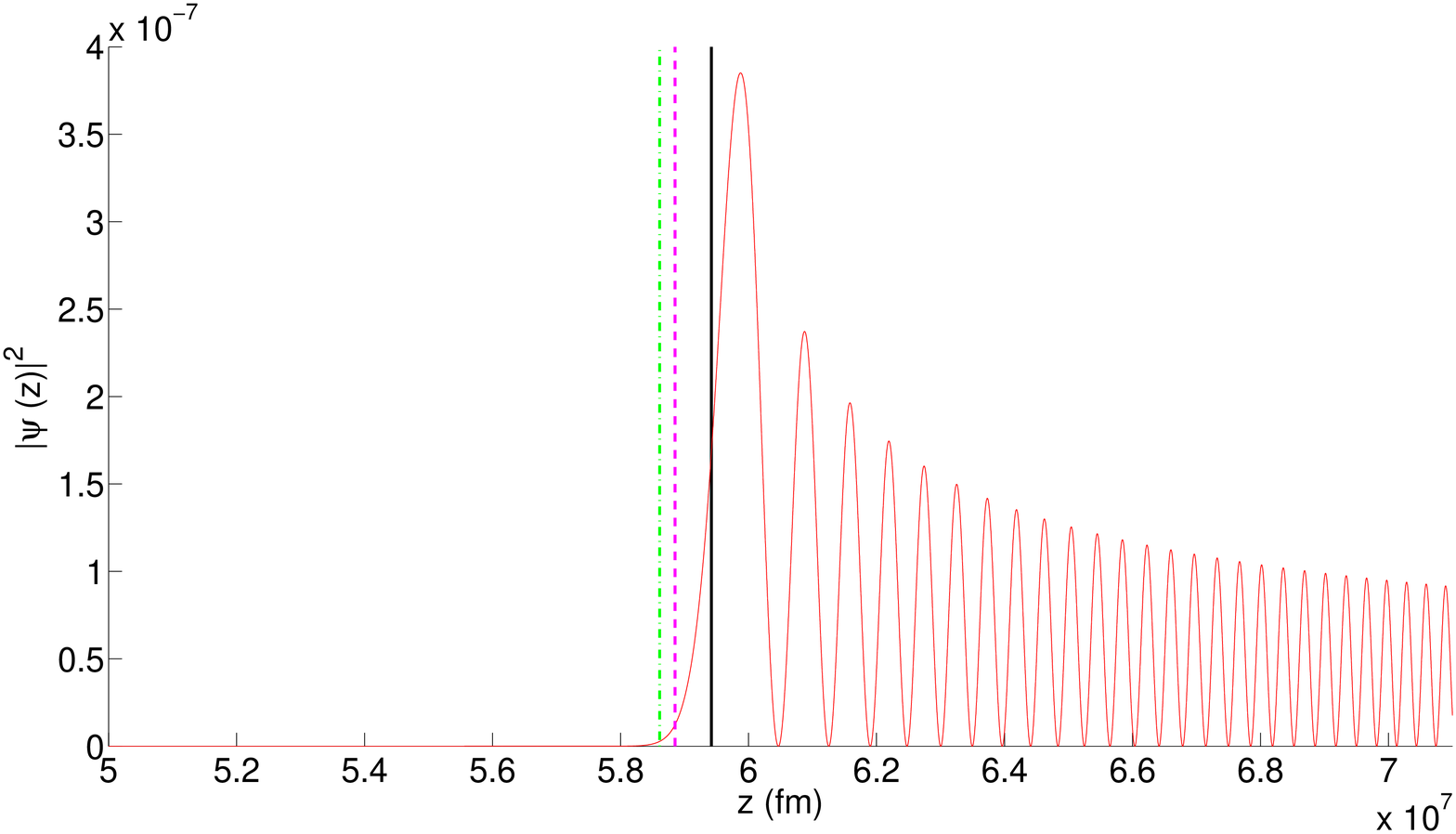}
     \includegraphics[width=90mm,height=50mm,clip]{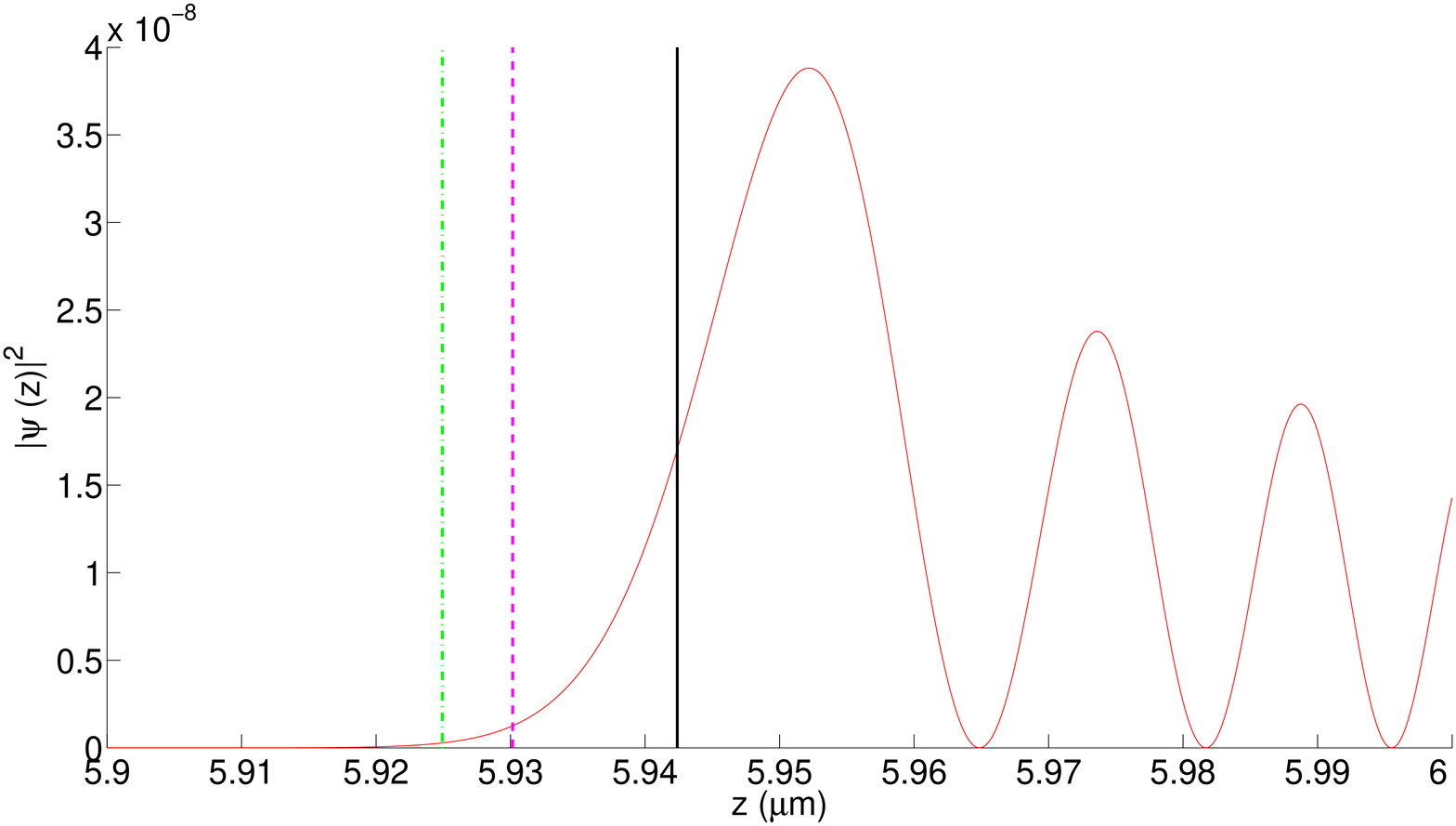}
     \end{center}
     \vspace{-0.5cm}
     \caption{Numerical solutions for the probability amplitude $|\psi|^2$, of electrons approaching the surface a CCO with an initial relative speed $\beta_{e^-}^{\rm ini.}$, equal to $10^{-2}$ (top) and $10^{-3}$ (bottom). The black (solid) vertical line corresponds to the classical minimum altitude $z_{\rm min.}^{\rm class.}$, the magenta (dashed) and green (dot-dashed) vertical lines correspond to the altitudes $z^{\rm 90\%}$ and $z^{\rm 95\%}$ respectively, defined such that 90\% and 95\% of the probability amplitude for $0<z<z_{\rm min.}^{\rm class.}$ lies at {\em larger} altitudes. The probability amplitude is normalised so that its sum for $z<z_{\rm min.}^{\rm class.}$ is equal to unity.}
     \label{tunneling}
     \vspace{-0.3cm}
\end{figure}

We also need to ensure that {\em quantum tunneling} effects do not significantly increase the rate of interactions between positrons and electrons, owing to electrons penetrating the classically-forbidden region of the positronsphere associated with significantly larger positron densities. To do this we use the Numerov method to numerically integrate the non-relativistic time-independent Schrodinger equation from within the classically forbidden region to larger altitudes. The results for the probability amplitude $|\psi|^2$, normalised so that the summed amplitude for altitudes below the classical minimum altitude $z_{\rm min.}^{\rm class.}$ is equal to unity, are displayed in figure~\ref{tunneling}.

We clearly observe that the majority of the probability amplitude within the classically-forbidden region (i.e. altitudes below the black (solid) vertical line) is extremely close to the classical minimum altitude, $z_{min.}^{class.}$ (95\% and 90\% of which is contained above the magenta (dashed) and green (dot-dashed) vertical lines located at altitudes $z^{95\%}$ and $z^{90\%}$ respectively). To demonstrate this quantitatively we calculate the following ratio

\begin{equation}
R=\frac{{\bar n_{e^+}}}{n_{e^+}(z_{min.}^{class.})}=\frac{\int^{z_{min.}^{class.}}_{z=0}|\psi(z)|^2n_{e^+}(z)dz}{n_{e^+}(z_{min.}^{class.})\int^{z_{min.}^{class.}}_{z=0}|\psi(z)|^2dz}>1.
\end{equation}

We find that $R=1.0104$ for $\beta=10^{-2}$, and $R=1.0022$ for $\beta=10^{-3}$, indicating that the 
increased density experienced by the tunneling electrons, relative to that experienced by purely classical electrons, is suppressed to within 1 percent of the density evaluated at the classical barrier $z=z_{\rm min.}^{\rm class.}$. Hence, we consider this to be ample justification for adopting the classical trajectories of electrons incident upon CCOs when calculating their rates of interaction with positrons residing in their positronspheres.

Because of the rapidly decreasing nature of the positron density with increasing altitude (as demonstrated by eq.~(\ref{n_p}) ), we expect that the probability for forming positronium is significantly reduced relative to when the incident electrons are allowed to traverse unhindered to the CCO surface. The value of this relative suppression $P$, will be a function of the CCO surface potential $V_c$, aswell as the initial speed $\beta_{e^-}^{\rm{ini.}}$ and impact parameter $b$ of the incident electron, relative to the CCO. The suppression factor $P$ can be expressed as

\begin{eqnarray}\label{P}
P={\rm exp}\left(-\int^{\infty}_{t=0}n_{e^+}(z[t])~\sigma_{e^-e^+\rightarrow P_s}(v_{\rm{rel}})~v_{\rm rel}~{\rm d}t\right).\nonumber\\
\end{eqnarray}

\noindent We can obtain an approximate {\em lower} limit on $P$ by adopting a radial electron trajectory, which approximately maximises the path integral of positrons, and by using a value of the parapositronium cross-section of $\sigma_{e^-e^+\rightarrow P_s}(v_{\rm{rel}})v_{\rm{rel}}\sim\pi a_0^2~\beta_{e^-}^{\rm{ini.}}c$, which is a generous upper estimate (see e.g. \cite{Igarashi} for the relationship between $\sigma_{e^-e^+\rightarrow P_s}$ and $\beta_{e^-}^{\rm{ini.}}$).

Using the above method, we obtain $P\le 10^{-3}$ for $\beta_{e^-}^{\rm{ ini.}}=10^{-2}$ and $P\le 10^{-7}$ for $\beta_{e^-}^{\rm{ ini.}}=10^{-3}$. Therefore, even if we adopt the estimate for the 511~keV flux $\Phi_{511}$ from positronium formation proposed by \cite{Oaknin}

\begin{equation}
\Phi\simeq 10^{-3}~{\rm cm}^{-2}~{\rm s}^{-1}\left(\frac{10^{33}}{B_{\rm CCO}}\right)^{1/3},
\end{equation} 

\noindent which is of the same order as that observed by the INTEGRAL satellite, if we now take into account the suppression on this flux dscribed above by multiplying it by $P$, we obtain

\begin{equation}
\Phi_{511}=10^{-3}~{\rm cm}^{-2}~{\rm s}^{-1}\left(\frac{10^{33}}{B_{\rm CCO}}\right)^{1/3} P\,\,,\nonumber\\
\end{equation} 

\noindent hence,

\[\Phi_{511}<\left\{ \begin{array}
                                                                               {r@{\quad{\rm for}\quad}l}
                                                                               10^{-6}~{\rm cm}^{-2}~{\rm s}^{-1}\left(\frac{10^{33}}{B_{\rm CCO}}\right)^{1/3} & \beta_{e^-}^{\rm ini.}=10^{-2} \\ 
                                                                               10^{-9}~{\rm cm}^{-2}~{\rm s}^{-1}\left(\frac{10^{33}}{B_{\rm CCO}}\right)^{1/3} & \beta_{e^-}^{\rm ini.}=10^{-3}\,\,,
                                                                                   \end{array} \right.     \] 

\noindent which is consistent with the INTEGRAL observations for $B_{\rm CCO}<10^{22}$ for $\beta_{e^-}^{\rm{ini.}}=10^{-2}$
\footnote{
However we should like to point out that such speeds for the majority of electrons traversing the ISM are optimistically high in conventional models describing the rotational velocity profile of the galaxy.
}
, but not reconcilable for electrons incident with speed $\beta_{e^-}^{\rm{ini.}}=10^{-3}$, even when using the fore-mentioned experimental lower limit $B_{\rm CCO}=10^{20}$.

We should also note that in much of the literature relating to strange stars composed of similar colour-superconducting matter (e.g. \cite{Alcock, Harko}), it is proclaimed that such stars possess positron/electrospheres which obey equations identical to eqs.~(\ref{V})-(\ref{n_p}), but are only several 1000 fm in depth, compared to the infinitely extending atmospheres adopted in the present study. If we were to adopt such truncated positronspheres, the charge associated with the bare antiquark matter would not be entirely shielded, yielding CCOs (plus positronspheres) with a net electrical charge. Even if this charge is small enough to be consisent with observations relating to charged dark matter, the CCOs will still possess electric fields which repel incident electrons as discussed above. Furthermore because of the reduced screening associated with these truncated positronspheres these fields will be {\em more} intense at large, positron-deficient altitudes. Thus not only will the incident electrons be repelled at altitudes {\em larger} than those calculated above, based on the above calculations they will be repelled far {\em above} the truncated positronsphere, and the interaction rate with positrons will be significantly {\em less} than that calculated above, yielding a significantly {\em smaller} value of the suppression factor $P$. Hence, the above scenario should be considered to be `optimistic' as far as explaining the 511~keV line signal with CCOs is concerned, and we will continue to adopt this scenario throughout.

It has also been proposed that the excess of gamma-rays detected by COMPTEL at energies $\simeq1-20$~MeV \cite{Strong} can be naturally explained by the photons produced in the (non-resonant) direct annihilation process involving interstellar electrons and positrons surrounding CCOs \cite{Lawson}. The appeal of such an explanation is that the energy at which the excess occurs is similar to the Fermi energy of the positrons located at the CCO surface, and hence direct annihilations involving such positrons would yield photons which may potentially contribute to this excess.

However, in \cite{Lawson} it is assumed that the incident electrons can penetrate the positronsphere ``unhindered'', right down to the surface of the CCO, and directly annihilate with positrons whose Fermi energy (which at any altitude is equal to the electric potential energy $V(z)$, see eq.(\ref{V}) ) is approximately $V_{\rm c}\simeq$\,20~MeV. From the above discussion we see that the penetration of interstellar electrons to altitudes below the classical minimum altitude $z_{\rm min.}^{\rm class.}$, which is of order $10^7$~fm for $\beta_{e^-}^{\rm ini.}=10^{-2}$ and $10^9$~fm for $\beta_{e^-}^{\rm ini.}=10^{-3}$, is exponentially suppressed due to the rapidly increasing electric potential as it approaches CCO surface.

Consequently, the electrons will only be able to directly annihilate with positrons at $z>z_{min.}^{class.}$ with Fermi energies $\mu_F(z<z_{\rm min.}^{\rm class.})\le V(z_{\rm min.}^{\rm class.})=T_{e^-}^{\rm ini.}$, where $T_{e^-}^{\rm ini.}$ is the kinetic energy of the incident electron, and equal to $\simeq3\times10^{-5}$~MeV for $\beta_{e^-}^{\rm ini.}=10^{-2}$ and $\simeq3\times10^{-7}$~MeV for $\beta_{e^-}^{\rm ini.}=10^{-2}$, i.e. in both cases $\ll~1~$MeV.

There is a finite probability that such electrons may quantum tunnel through the classical potential barrier to the surface of the CCO, and here we estimate an upper limit for which as follows.
The solution for the electron probability amplitude $|\psi(z)^2|$ locally around a given altitude $z<z_{min.}^{class.}$ is of the form

\begin{equation}
|\psi(z)|^2\propto{\rm exp}[\kappa(z) z],\quad\kappa(z)=\left[\frac{2m\left(V(z)-T\right)}{\hbar}\right]^{1/2}.
\end{equation}

\noindent Since $V$ monotonically increases with decreasing $z$, the value of the exponent $\kappa$ also increases with decreasing $z$, so that we may say

\begin{equation}
\frac{|\psi(0)|^2}{|\psi(z_{\rm min.}^{\rm class.})|^2}<\frac{|\psi(z'>z_{\rm min.}^{\rm class.})|^2}{|\psi(z'>z_{\rm min.}^{\rm class.})|^2}{\rm exp}[-z'\kappa(z')]\,\,,
\end{equation}

\noindent where the LHS is the probability of an incident electron located at $z_{\rm min.}^{\rm class.}$ to quantum tunnel to the CCO surface, and the RHS is obtained by extrapolating the probability density at $z'$ using a fixed exponent. Here for simplicity we use $z'=z^{99\%}$ and obtain

\[{\rm ln}\left(\frac{|\psi(0)|^2}{|\psi(z_{\rm min.}^{\rm class.})|^2}\right)<\left\{ \begin{array}
                                                                               {r@{\quad{\rm for}\quad}l}
                                                                               -388 & \beta_{e^-}^{\rm ini.}=10^{-2} \\ 
                                                                               -1808 & \beta_{e^-}^{\rm ini.}=10^{-3}\,\,,
                                                                                   \end{array} \right.     \] 

\noindent which effectively means that the probability for an electron to tunnel to positrons which possess Fermi energies large enough for their direct annihilations to contribute to the MeV gamma-ray excess is vanishingly small.
 
However, electrons resulting from pair production, due to the extremely strong electric fields near to the CCO surface \cite{pair_emission, Harko}, may directly annihilate with positrons in the surrounding positronsphere at sufficiently small altitudes in order to produce MeV gamma rays which may contribute to the MeV excess. To calculate such fluxes requires a knowledge of the thermal structure of each CCO which lies outside of the scope of this study 
  
We should also mention that positronium atoms which may form  close to the CCO surface involving electrons either resulting from pair production, or those undergoing extremely improbable tunneling events, would likely be quickly reionized by the intense static electric field originating from the bulk anti-quark matter.
We expect that positronium atoms would quickly destabilize or have its initial formation suppressed when a static electric field exceeding strengths of order $13.6~$eV~\AA$^{-1}$ is present. Substituting $V_c=20~$MeV into eq.\,(\ref{E}), we observe that $E<13.6~$eV\AA~$^{-1}$ for $z\ge3.34~$nm. Hence, we expect that Positronium formation will be suppressed inside this radius due to this effect, which 
for tunneling electrons will be in addition to the extreme levels of suppression discussed above.\\

\noindent {\em Conclusions:-} In this study we have considered the possibility of compact composite object dark matter, composed of colour-superconducting di-antiquark pairs, as a solution to several unexplained astronomical observations. Firstly, we considered the 511~keV line signal emerging from the galactic centre and observed by INTEGRAL, and secondly the excess of background gamma-rays with energies approximately in the range 1-20~MeV, and observed by COMPTEL.

It was proposed that the 511~keV excess could be explained by the decay of parapositronium atoms formed from interstellar electrons and positrons electrostatically bound in a positronsphere surrounding each CCO. It was also proposed that the COMPTEL MeV excess could be produced from the direct annihilation of interstellar electrons and high energy positrons located near to the surface of the bulk matter of the CCO. 

In such proposals it was assumed that the incident electrons could travel unhindered to the CCO surface, however in this study we have deduced that insterstellar electrons approaching these CCOs with typical speeds will be unable to reach altitudes below approximately $10^7$~fm (compared to zero). Consequently, the rates of positronium formation and the rate of direct annihilation at the CCO surface are significantly reduced; in the case of positronium formation the relative suppression $P<10^{-3}$, and the probability for direct annihilation at the CCO surface is exponentially suppressed to zero since such annihilations can only occur by incident electrons which tunnel through the potential barrier. 

Hence, in conclusion we find that the respective contributions of these processes to the fore-mentioned excesses are significantly reduced relative to previous estimates, to the extent where a CCO explanation of INTEGRAL and COMPTEL excesses is significantly less motivated.\\

\noindent DTC is supported by the Science and Technology Facilities Council. GDS thanks the Beecroft Institute at Oxford for hospitality and for support during
parts  of this research.  GDS's research was also supported in part by the Guggenheim Foundation,
and by the US DoE.

\end{document}